\documentclass[%
 reprint,
 amsmath,amssymb,
 aps,
]{revtex4-1}

\usepackage{graphicx}
\usepackage{subcaption}
\usepackage{dcolumn}
\usepackage{bm}
\usepackage[colorlinks=true,linkcolor=red,urlcolor=blue,citecolor=red]%
		{hyperref}

\newcommand{\reffig}[1]{Fig.~\ref{#1}}
\newcommand{\refeq}[1]{Eq.~\ref{#1}}

\newcommand{\refsec}[1]{Sec.~\ref{#1}}
\newcommand{\refcite}[1]{Ref~\cite{#1}}

\begin{document}

\preprint{APS/123-QED}

\title{The Dose of the Threat Makes the Resistance for Cooperation}

\author{Uzay Cetin}
  \email{uzay00@gmail.com}
\affiliation{Department of Computer Engineering, Bogazici University.}
\author{Haluk O. Bingol}%
\affiliation{Department of Computer Engineering, Bogazici University.}

\date{\today}

\begin{abstract}
We propose to reformulate the payoff matrix structure of 
Prisoner's Dilemma Game,
by introducing threat
and greed factors, and show their effect on the co-evolution of memory and
cooperation. 
 Our findings are as follows. 
(i)~Memory protects cooperation. 
(ii)~To our surprise,
greater memory size is unfavorable to evolutionary  success when there is no threat. In the absence of threat, subsequent generations lose their memory and are consequently invaded by defectors. 
(iii)~In contrast, the presence of an appropriate level of threat triggers the emergence of 
 a self-protection mechanism for cooperation,
which manifests itself as an increase in memory size within subsequent generations. 
 On the evolutionary level, memory size acts like an immune response of the generations against aggressive defection. 
 (iv)~Even more extreme threat results again in defection. 
 Our findings boil down to the following: The dose of the threat makes the resistance for cooperation.
\end{abstract}

\maketitle


\section{Introduction}
Taking cooperative actions against a common threat, is frequently seen in nature and in history as well. 
Herbert Spencer puts it as follows, 
``Only by imperative need for combination in war were primitive men led into cooperation"~\cite{Spencer}. 
Individuals, as a response to what they perceive as threat, bind together and tend to move as a unit. 
Similar collective spirit, 
can also be seen in fish swimming in schools or birds flying in flocks. 
The waves of agitation in schools or flocks are nothing but an escape maneuver from an attack of a predator~\cite{CharlotteK2015Article}. 
Kin selection, direct reciprocity, indirect reciprocity, group selection and limited local interactions 
are shown to be five powerful determinants of cooperation~\cite{Nowak08122006, rand2013human}. 
Yet, explaining cooperation still remains one of the greatest challenges across disciplines~\cite{Pennisi01072005}.
Here, we discuss the dose of the threat imposed by environment as another way to obtain cooperation.

In Ref~\cite{%
	Wright2000Book},
Robert Wright says, ``interaction among individual genes, or cells, or animals, among interest groups, 
or nations, or corporations, can be viewed through the lenses of game theory".
Nevertheless, the amount of information stemming from the huge number of interactions, 
can easily exceed the processing capabilities of the interacting parties. 
This is also referred as attention scarcity problem in the literature~\cite{%
	Falkinger2008,%
	Cetin2014b}.
In our previous work, we coined the term  \emph{Attention Game} to define an interacting environment 
where players can only pay attention to a portion of the information they receive~\cite{%
	Cetin2014a}.
We worked on attention games in a specific context of Iterated Prisoner's Dilemma (\emph{IPD}).

\begin{figure*}[!htbp]
	\begin{subfigure}{0.50\columnwidth}%
	\centering
		\includegraphics
			{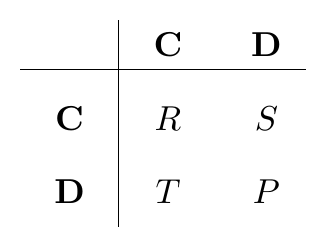}
		\caption{Generic PD} 
		\label{fig:payoffMatrixSPRT}
	\end{subfigure}
	\hspace*{0.8cm}
	\begin{subfigure}{1\columnwidth}%
		\centering
		\includegraphics[scale=0.8]
			{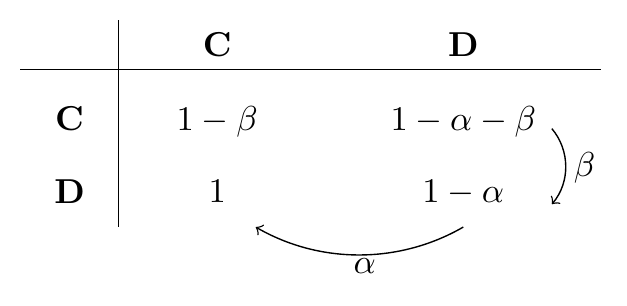}
		\caption{
		Threat Game
		} 
		\label{fig:payoffMatrixAlphaBetaNormalized}
	\end{subfigure}
	~\\
	\begin{subfigure}{0.50\columnwidth}%
	\centering
		\includegraphics
			{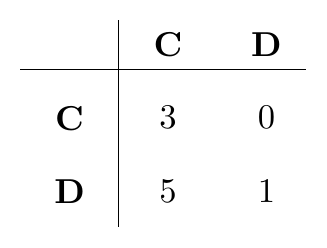}
		\caption{Axelrod} 
		\label{fig:payoffMatrixAxelrod}
	\end{subfigure}
	~
	\hspace*{0.8cm}
	\begin{subfigure}{0.50\columnwidth}%
	\centering
		\includegraphics[scale=0.8]
			{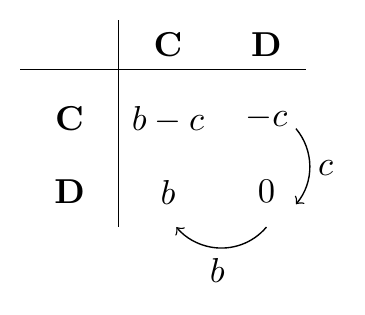}
		\caption{Donation Game} 
		\label{fig:payoffMatrixDonation}
	\end{subfigure}
	\begin{subfigure}{0.50\columnwidth}%
	\centering
		\includegraphics[scale=0.8]
			{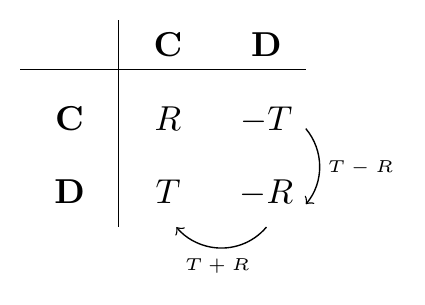}
		\caption{Epstein} 
		\label{fig:payoffMatrixEpstein}
	\end{subfigure}
	\caption{
		Payoff matrices.
	} 
	\label{fig:payoffMatrices}
\end{figure*}

Prisoner's Dilemma (\emph{PD}) game is a standard model,
described by the payoff matrix shown in \reffig{fig:payoffMatrixSPRT},
to study how selfish beings manage to cooperate~\cite{Axelrod1984Book}.
There are two types of actions, namely,
cooperation (\emph{C}) and defection (\emph{D}).
Two agents select their actions to play against each other.
In the case of mutual cooperation, both receive the \emph{reward} payoff $R$. 
If one cooperates while the other defects, 
cooperator gets the \emph{sucker} payoff $S$ 
and defector gets the \emph{temptation} payoff $T$. 
In the case of mutual defection, 
both get the \emph{punishment} payoff  $P$. 
To be an Iterated Prisoner's Dilemma Game, 
the following conditions must hold for the payoffs: 
$S < P < R< T$ and $T + S<2R$~\cite{Axelrod1984Book}.

Evolutionary game theory applies mathematical and computational techniques to study 
the evolution of cooperation. 
For an important review on co-evolutionary processes, see \cite{Perc2010109}.
It is shown that choice and refusal of partners 
accelerates the emergence of cooperation~\cite{%
	Tesfatsion1993}.
Memory is a prerequisite for engaging in reciprocity
and also for partner selection on the basis of past encounters.
In memory-based Prisoner's Dilemma Games, 
each player 
can keep track of only a limited number of the previous rounds
for all of its partners~\cite{MemoryBasedIPD}.
This limited number is defined as the 
\emph{memory length}~\cite{alonso2001effect, hauert1997effects}. 
Tit-for-Tat, the winner of the Axelrod's tournament, is a memory-one strategy. 
It starts with cooperation and afterwards imitates the last action of its partners~\cite{%
	Axelrod1984Book}.
Thus the memory length of agents using Tit-for-Tat is one,
even though they keep track of all of their partners.
Dunbar's number indicates a cognitive limit to the number of individuals with whom one can maintain stable relationships~\cite{%
	Dunbar1993}.
We think that the ability to keep track of all potential game partners is not always possible.
This may be thought as a natural consequence of huge amount of game partners 
or a very limited memory size to be informed of all.
The concept of memory in Prisoner's Dilemma, is generally explored in terms of 
historical time-dependency of previous rounds~\cite{szolnoki2009impact, szolnoki2010dynamically, Zhang2015}. 
Differently, from our perspective, the term \emph{memory size} indicates 
	the number of potential game partners one can keep track of.
	In our model, agents store a very brief information about the general behavior of a limited number of their partners.
	This information will be used to distinguish defectors from cooperators.

Evolutionary psychologists demonstrated that social exchange in  a group 
requires the existence of some mechanisms for detecting cheaters, 
but do not require any mechanisms for detecting altruists~\cite{%
	Cosmides1992Book}.
Similarly,
in our previous work
we found that
it is crucial for attention to be focused on defectors in order to foster cooperation 
when agents have insufficient memory size~\cite{%
	Cetin2014a}.
That is, 
attention should be allocated in such a way that
agents should keep remembering defectors, and forget preferentially cooperators 
whenever memory is exhausted.
In Ref~\cite{%
	Cetin2014a}, 
memory size does not differ from one player to another and there exists only two type of players such as pure cooperators and pure defectors.
We will use a similar attention mechanism focusing on defectors.
In this study, we will introduce heterogeneity to our work
by allowing agents to have different memory sizes and strategies.
We will investigate how the characteristics of agents evolve from generation to generation.

The essence of how selfish beings manage to cooperate is captured 
by the payoff matrix of the IPD game.
Axelrod used the fixed payoff matrix
given in \reffig{fig:payoffMatrixAxelrod}
for his tournament~\cite{%
	Axelrod1984Book}. 
A natural extension would be to investigate the impact of different payoff entries.
Many studies use payoff matrices with positive payoffs.
Some works on negative payoffs are also done.
To this end, 
some researchers prefer to fix the two selected entries of the payoff matrix 
and explore the effect of the change in the other two entries~\cite{%
	Epstein,%
	Nowak08122006,%
	Nowak2015}. 
In the so-called \emph{Donation game},
given in \reffig{fig:payoffMatrixDonation},
cooperation corresponds to offering the other player a benefit $b$ at a personal cost $c$ 
and defection corresponds to offering nothing.
Nowak has investigated the effect of these two essential parameters 
for various situations~\cite{%
	Nowak08122006}.
This is a very agreeable representation, 
but it requires $P = 0$.
So it does not allow to study the dynamics of cooperation where 
the punishment payoff $P$ is positive or negative.
In Ref~\cite{%
	Epstein},
Epstein investigated the case of negative sucker and punishment payoffs
for a special case of $S = -T$ and $P = -R$,
given in \reffig{fig:payoffMatrixEpstein}.
	We also want to study the case where receiving a defection leads to negative payoffs.
	Differently, we interpret the case of $S<P<0$ as the presence of threat.
	For $S<P<0$,
	the decrease in the negative values of $S$ and $P$ corresponds to 
	the increasing level of threat.
	From our perspective,
	non-negative values of $S$ and $P$ corresponds to the absence of threat.
To investigate the effect of threat,
we propose to use a more general parametric payoff matrices of the form \reffig{fig:payoffMatrixAlphaBetaNormalized}
which covers the family of payoff matrices of the form given in  \reffig{fig:payoffMatrixDonation} and 
\reffig{fig:payoffMatrixEpstein}.

This article is structured as follows:
In the next section 
we explain our motivation 
and
in \refsec{sec:Model}
we present our agent-based model 
and give the technical details of the simulations for a generic payoff matrix. 
In \refsec{sec:Results}, 
we provide the results for two specific payoff matrices: 
(i)~one with all non-negative entries,
and
(ii)~the other with negative entries for sucker and punishment payoffs.
In \refsec{sec:Payoff}, 
we generalized the payoff matrices with two parameters.
Finally, in \refsec{sec:Conclusion} 
we summarize our findings and
construct some analogies with various disciplines.

\section{Motivation}
\label{sec:Motivation}

Consider selfish agents playing evolutionary IPD game.
Assume that the fitness of agent is correlated with its accumulated payoffs.
Then, in order to increase its fitness, 
a selfish agent tries to maximize its gain at every single round of the game.
Suppose agents have the right to choose or refuse to play.
If all the entries of the payoff matrix are non-negative,
should an agent choose to play with every opponent whether it is a defector or not?

Agents with myopic view may prefer immediate positive outcomes in the short-term at the expense of longer-term outcomes.
When interacting with an opponent brings relatively low payoff,
the agent will accumulate less payoffs compared to that opponent
and at the end, 
the agent will have a lower chance to reproduce. 
This is the case of cooperator playing against defector.  
If only cooperators could have find a way to distinguish defectors from cooperators and refuse 
to play with the defectors, then 
the cooperators can outcompete 
the isolated defectors. 
So the macro-level dynamics of the population depends on the
mixture of agents: how cooperative and with whom willing to play they are.

It is not possible for cooperation to flourish in a well-mixed population
without any mechanism that give cooperators the ability to quarantine defectors. 
Spatial structure can promote cooperation
by introducing physical barriers against interaction with defectors.
Static networks lack the ability for modeling the dynamical interactions~\cite{santos2006cooperation}.
So, recent advances make emphasis on the co-evolution of strategy and environment~\cite{Perc2010109}.
In this study,
we follow a different path in order to promote cooperation.
In lieu of considering spatially structured population in physical space,
we will consider conceptually structured populations.
Agents will have mental representations of other agents and they 
will have the ability to choose with whom to interact.
Our proposition fits nicely to the research line of conditional strategies~\cite{szolnoki2012conditional, szolnoki2013effectiveness, szolnoki2016competition}.
In our model, 
agents interact with all except the ones that they perceive and remember as defectors.
Thus, memory plays the role of conceptual barriers for interaction
with defectors.
If we consider payoff matrices with negative values,
for $S$ and $P$,
then the dynamics may become more complicated
but the need for the refusal of defectors becomes more clear.
For a cooperator, 
there was a risk of not gaining ($S = 0$) but now losing points becomes also a possibility ($S < 0$). 
Whenever $P$ also becomes negative,
defectors also face the risk of losing points.
To identify the characteristic of the opponent 
and if it is defector not to play with it becomes an essential asset
especially when receiving a defection leads to negative payoffs. 
We will consider risk of losing points as \emph{threat}.
Hence payoff matrices with negative entries, 
for sucker and punishment payoffs,
are considered to be games with threat.

Our main research question, in this article, will be the following. What is the effect of increasing level of threat on the co-evolutionary dynamics of memory and cooperation?

\section{Model}
\label{sec:Model}

We propose an evolutionary game,
where generations do not overlap.
In our model, there are $N$ agents playing
IPD game within the generation.
At the end of certain number of rounds 
the generation is terminated, 
and
all the agents of the old generation are removed.
Before the old agents are removed,
they reproduce according to their fitnesses.
Roulette wheel selection is applied $N$ times to pick agents that will reproduce. 
Hence, the population size is kept constant at $N$.
We set $N=100$.

\subsection{Rounds}
Agents interact and try to increase their accumulated payoffs
by means of playing a modified IPD~\cite{%
	Axelrod1984Book}.
In each round, 
two agents are selected uniformly at random
and 
given the chance to play.
Each selected agent has to decide whether 
to choose or to refuse to play with the given opponent.
If at least one of them 
refuses to play,
no playing takes place and  the round is completed,
hence their scores do not change.
If both agents 
choose  to play,
then they play the usual Prisoner's Dilemma game.
Each agent selects its 
action 
of either cooperate or defect.
According to their joint actions, 
each collects its payoff based on the generic payoff matrix given in \reffig{fig:payoffMatrixSPRT}.
The payoff collected is added to the cumulative payoff, 
called \emph{score}, 
and
the round is completed.

We want every pair of agents to be selected on average $\tau$ times.
Therefore each generation lasts $\tau {N \choose 2}$ rounds.
We use $\tau = 30$.

\subsection{Choice and refusal of partners.}
\label{sec:misperception}
Agents have probabilistic behavior. 
An agent $i$ simply chooses to defect with probability $\rho_{i}$,
called \emph{defection rate},
or to cooperate with $1 - \rho_{i}$.
The defection rate of an agent is a property that never changes.

Choice and refusal to play 
is based on
the agent's subjective perception of the opponent 
as a defector or a cooperator.
The agent refuses to play with an opponent 
if the opponent is perceived as a defector.
In order to decide whether the opponent is defector, 
agent uses its memory.
The agent keeps track of the previous rounds with the opponents.
That is,
for every opponent, 
it keeps two numbers,
namely,
the total number of rounds played with the opponent and 
in how many of them the opponent has defected.
The ratio of 
the number of received defections to 
the number of total rounds 
is called \emph{perceived defection rate}.
The opponent is perceived as \emph{defector} 
if its perceived defection rate is fifty percent or more. 
Otherwise, 
the opponent is perceived as \emph{cooperator}.
As a third case,
if there is no history about the opponent,
it is considered as if it is cooperator. 
Namely, the default decision is to play.

We should give some intuition about agent's possible misperceptions 
due to the small number of interactions, at this point.
Different agents can perceive the same agent differently, at the same time.
Suppose agent $i$ has a low defection rate which is greater than zero.
Then, 
in most of the games it plays, 
it will cooperate
and in a very few of them it will defect. 
Therefore it is expected that most of the agents consider it as ``cooperator''.
But it is still possible that 
some agents can perceive it as a ``defector"
and refuse to play with it again.
This may happen 
due to the small number of interactions with $i$,
in which $i$ happened to 
defect more than cooperate.
In statistics, it is known that the small sample size 
may not be a good representative of a probability distribution.
But our agents do not know it, like most of the people~\cite{%
	Kahneman2011Book}.

\subsection{Memory}

Remembering the history of the opponents calls for memory for each agent.
Every agent has different size of memory.
Let $M_{i}$ denotes the number of opponents agent $i$ can keep track of.
The ratio $\mu_{i} = M_{i} / N$ is called the
\emph{memory ratio} 
 of agent $i$.
If $M_{i} \ge N$,
then agents have enough space to recall every agent.
Since this case is not interesting,
we investigate the case where $M_{i} \le N$ for all $i$.
That is, 
agents do not have enough memory space to store the history of all opponents.
Hence $\mu_{i} \in [0, 1]$.

If the number of rounds in a generation is big enough,
an agent encounters with almost all agents and
in order to keep the history of each opponent,
it requires memory size of $N$.
Suppose agents have limited memory size, 
i.e., $M < N$.
Then after $M$ different opponents,
there is no room left for the $M + 1^{th}$ opponent.
This requires a selective attention mechanism. 
Agents should decide 
which agents to keep in memory and which agent to forget.
Our previous study indicates that
it is a better strategy to focus on defectors rather then cooperators~\cite{%
	Cetin2014a}.
So if there are memory spaces reserved for cooperators,
select a cooperator.
If there is no cooperator left,
then select a defector.
Then forget the selected opponent and use this reclaimed space for the new opponent.
Both cooperator or defector selections are done by uniformly at random.
\subsection{Fitness}
In evolutionary games,
agents reproduce proportional to their fitness.
We define fitness as a function of scores.
Agents of a new generation start with zero scores. 
As they play, 
the payoffs obtained are added to the score.
In the traditional Prisoner's Dilemma game, 
the payoffs are all non-negative 
such as $(S, P, R, T) = (0, 1, 3, 5)$. 
Hence playing will not decreases the score.
In this study we also consider payoff matrices with negative entries, too.
In the case of negative payoffs, 
the scores of agents may decrease 
and negative scores are possible.
Therefore, using scores directly as fitnesses will not work.
The mapping the scores to fitness values requires attention.
We adjust the scores 
by subtracting the minimum score from all.
After this, adjusted scores become all non-negative. 
Then obtain the normalize score by 
dividing the adjusted scores to 
the sum of all adjusted scores.
Then, use the normalized  
scores as the fitness for reproduction.
Note that 
the agent with the minimum score has
the normalized score of zero.
Hence it does not reproduce.

\subsection{Reproduction}

The reproduction is asexual. 
Each child has exactly one parent.
The \emph{genotype} of  an agent $i$, is the pair $(\mu_i, \rho_i)$.
A child gets the exact copy of the genotype of its parent if there is no mutation.
With probability of $r$,
called \emph{mutation rate},
there is a mutation.
When there is a mutation, 
only one of the entries, 
selected at random, 
in the genotype is replaced with 
a new number drawed from a uniform distribution of $[0, 1]$.
We use $r = 0.05$.

\subsection{Visualization}

Note that there is a useful visualization for genotype $(\mu_i, \rho_i)$.
Agents can be represented by points on 
the unit square of the $\mu$-$\rho$ plane 
where $x$-axis is the memory ratio $\mu$ and 
$y$-axis is the defection rate $\rho$.
The point $(\mu_i,\rho_i)$ displays the genotype of the $i$'th agent.
The \emph{average defection rate} and the \emph{average memory ratio} of the current population are given by 
$\overline{\mu} = \frac{1}{N} \sum^N_{i=1} \mu_i$ 
and 
$\overline{\rho} = \frac{1}{N} \sum^N_{i=1} \rho_i$, respectively.
We can picture the \emph{average genotype} ($\overline{\mu}$,$\overline{\rho}$), 
as a point on that phase plane as in \reffig{fig:PhasePlaneAxelrod}.

\subsection{Initialization and termination}

Once an initial generation is formed,
system runs from one generation to the next with the given dynamics.
The parameters of the agents of the initial generation are set randomly using uniform distributions.
That is,
for each agent $i$
the values for the genes 
$\mu_i$ and $\rho_i$ are set using a uniform distribution over $[0, 1]$.
The number of generations, 
before the simulations are terminated,
is another model parameter.
We terminate our simulations after 500 generations.

\section{Results}
\label{sec:Results}

\begin{figure}[h]
	\begin{center}
		\includegraphics[scale=0.45]{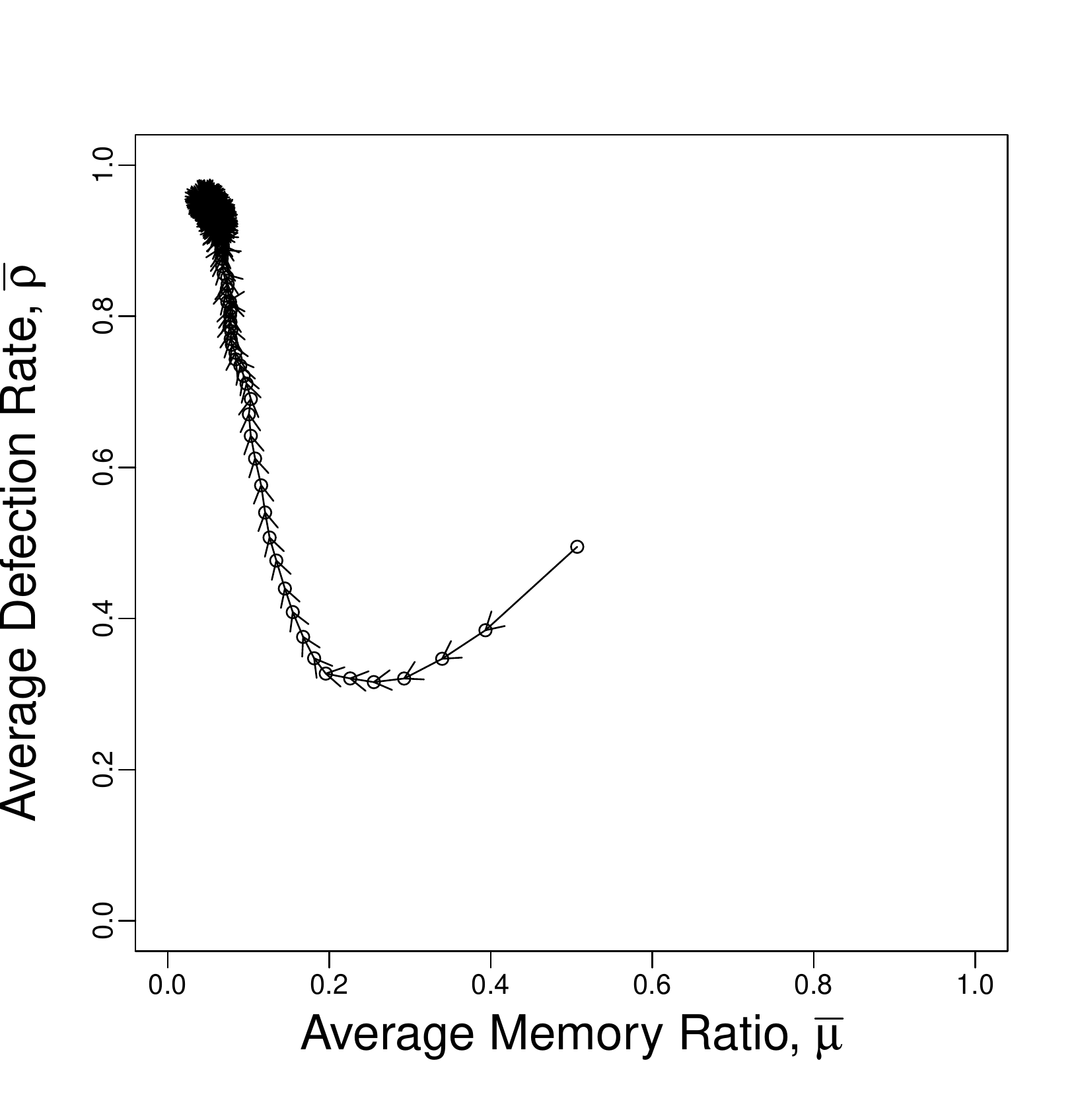}
	\end{center}
	\caption{
		Co-evolution of average memory ratio $\bar{\mu}$ and average defection rate $\bar{\rho}$ through generations. 
		$(S,P,R,T) = (0,1,3,5)$.
	}
	\label{fig:PhasePlaneAxelrod}
\end{figure}

We run simulations using various payoff matrices.
Initial population starts with an average genotype close to $(0.5, 0.5)$. 
Tracking the values of 
$( \overline{\mu}, \overline{\rho})$
pairs from generation to generation will make us see the co-evolution of cooperation and memory, as in \reffig{fig:PhasePlaneAxelrod}.

\subsection{Absence of threat}

We have discussed that payoff matrices with negative entries cause threat to the agents.
We refer the case of non-negative payoffs, 
i.e., $0 \leq S < P < R < T$, 
as \emph{absence of threat}.
In the first set of simulations, 
we used the standard payoff values of $(S,P,R,T) = (0,1,3,5)$.
An averaged trajectory over $50$ different realizations of the same initial population 
can be seen in \reffig{fig:PhasePlaneAxelrod}.
Two dynamics are observed. 
(i)~Average memory size tends to decrease independent of the average defection rate of the population. 
Neither cooperation nor defection favor greater memory size when there is no threat.
(ii)~Average defection rate decreases if memory size is high and increases if it is low. 
Average defection rate $\overline{\rho}$ decreases at the beginning since 
initial value of $\overline{\mu} = 0.5$ is relatively high.
Interestingly, there is an unconditional decrease for $\overline{\mu}$. 
Once average memory ratio $\overline{\mu}$, becomes small enough,
average defection rate $\overline{\rho}$ starts to increase.
Memory size has a negative effect on defection rate,
in other words, memory protects cooperation.
Without memory,
i.e., ($\overline{\mu} \rightarrow 0$), 
cooperation becomes vulnerable and defection succeeds,
i.e., ($\overline{\rho} \rightarrow 1$).
The average genotype of the population gradually reaches to a point very close to $(0, 1)$. 
That is, agents become memoryless and defective when there is no threat.

It is known that evolution may lead to unexpected paths. 
The observation of the unconditional decrease of the memory size is totally unexpected. 
To understand it, first consider a population that is composed of defectors only. 
Is it better for defectors to have greater memory size? 
The answer is no, as long as punishment payoff $P$ is greater than zero. 
The reason is as follows: defectors with high memory size lose punishment payoff $P > 0$,
just because they remember and refuse other defectors. 
Thus they end up with lower fitness and they are eliminated throughout the evolution.  
Consider the second extreme case where the population is composed of cooperators only.
This case is a bit trickier. Previously we determined how agents perceive the world. 
Perception is open to mistakes 
as it is the case for real life. 
A cooperator with a low defection rate can be perceived as a defector, 
just because it is happened to defect more than cooperate 
within a small number of interactions.
As a result of this misperception, high memory size can cause to avoid engaging rounds with agents whose intention is mostly cooperate. 
Cooperators with high memory size end up with lower fitnesses.  
The relative abundance of the cooperators with high memory in the subsequent generations decreases,
and the relative abundance of the cooperators with low memory increases.
This manifests itself as a reduction in the average memory size, $\overline{\mu}$.

The surprising downside of having a greater memory size is isolation, which leads to a deficient fitness. 
Thus, by means of mutations, subsequent generations get rid of their memory 
in the absence of threat.
Without memory, defectors invade the subsequent generations.

\subsection{Presence of threat}

We investigate the outcomes of
an alternative formulation of negative payoffs, as in Ref~\cite{%
	Epstein}.
For $S = 0$, refusing or playing with a defector 
is apparently indifferent for a cooperator. 
Once $S$ becomes negative,
picture changes.
From the perspective of defectors, 
it is still better to play 
whatever the opponent type is,
as long as $S < 0 \le P$.
When $P$ becomes negative,
defectors have to be careful, too.
It is known that evolution is about the survival of the most suited organisms for the current environmental conditions. 
When we use a different payoff matrix, environment differs and dynamics dramatically change. 
Let's define \emph{aggressive defection}
as an harmful act that reduces the score of the agents that are subjected to it. 
PD game under aggressive defection can be given 
with an additional constraint of $S < P < 0$. 
Now, receiving a defection results in negative values and it hurts. 
Thus having a greater memory size may become advantageous, 
in contrast to the case of non-negative payoff matrix. 

\begin{figure*}[t]
	\hspace*{-0.5cm}
		\includegraphics[scale=0.35]{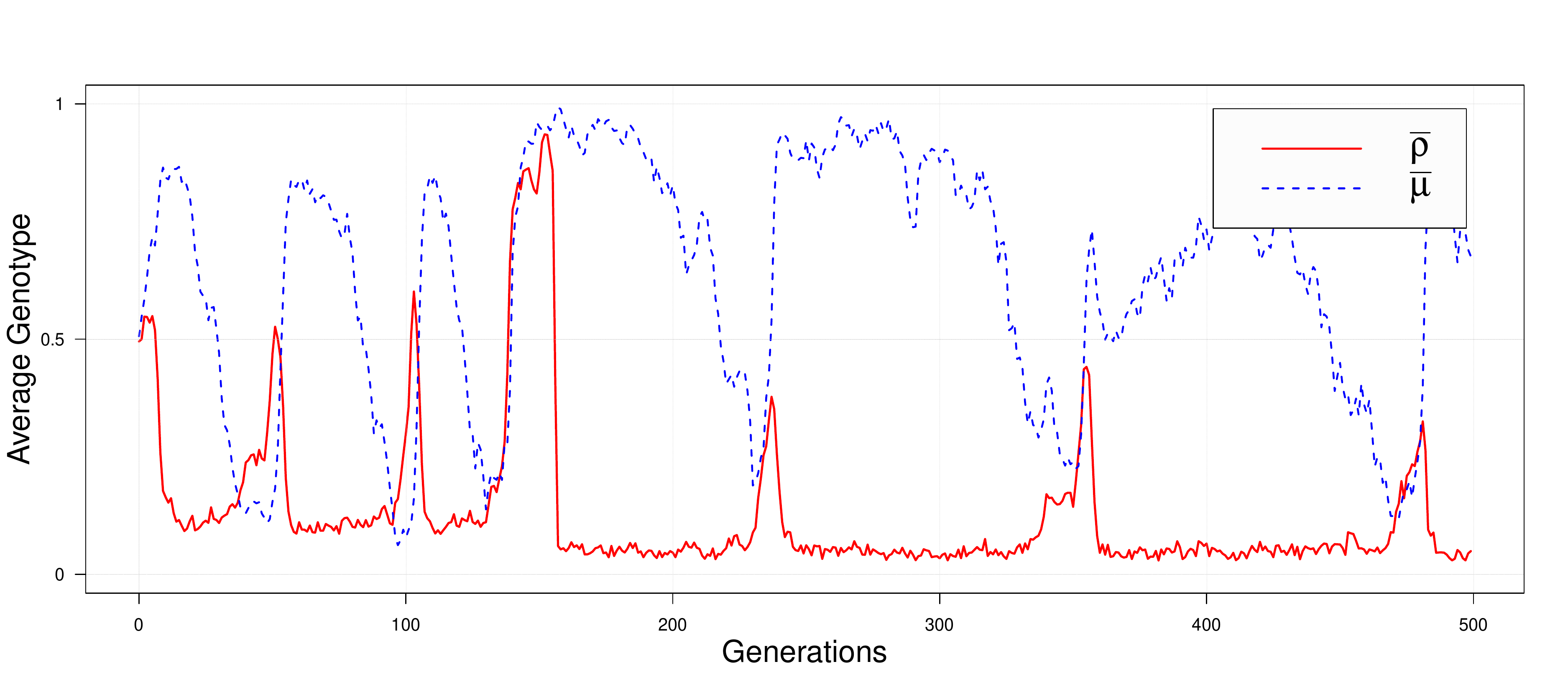}	
	\caption{
		Co-evolution under aggressive defection for a single realization.
		$x$-axis represents the generation steps 
		and $y$-axis represents the average genotype of the population.		 
		$(S,P,R,T) = (-7,-6,4,5)$.
	}
	\label{fig:harmfulA}
\end{figure*}

In the presence of threat, two dynamics begin to compete at the evolutionary level. 
(i)~Tendency to increase memory size, in order to maintain self-protection when average defection rate gets higher. 
(ii)~Tendency to decrease memory size, to avoid self-isolation when average defection rate gets lower. 
These two dynamics can give rise to oscillatory behaviors. 
In \reffig{fig:harmfulA}, we display the dynamics of a single realization 
for a biased payoff matrix of $(S, P, R, T) = (-7, -6, 4, 5)$. 
At generation 0 in \reffig{fig:harmfulA},
simulation starts with a randomly generated initial population 
whose average memory ratio is relatively high,  $\overline{\mu} = 0.5$. 
Agents with high memory size can protect themselves from defection. 
Thus defectors incur isolation and their fitness diminishes, $\overline{\rho} \rightarrow 0$. 
Eventually, cooperators with high memory size fill the population. 
When almost all agents turn out to be cooperator, 
around generation 20 in  \reffig{fig:harmfulA},
misperception becomes an issue. 
High memory size may block interactions among cooperators and this is the reason why 
evolution prefers cooperators with smaller memory size, $\overline{\mu} \rightarrow 0$. 
Population without a valuable memory provides an excellent opportunity not to be missed by mutant defectors. 
Thus population starts to be filled by defectors and the average defection rate of the population increases
around generation 50 in \reffig{fig:harmfulA}. 
Only cooperators with high memory size can resist to defectors, 
let's call them \emph{skeptic cooperators}. 
If there exists still some critical number of skeptic cooperators, 
resistance can take place and defectors can be outcompeted. 
That is, cooperators with high memory size again fill the population,
as it is seen around generation 60. 
This cyclic behavior repeats itself 
until relative abundance of the skeptic cooperators becomes inadequate to resist defectors.
In that case, defectors can invade the population
as it is seen around generation 150.

\begin{figure}[h]
\centering
		\includegraphics[scale=0.45]{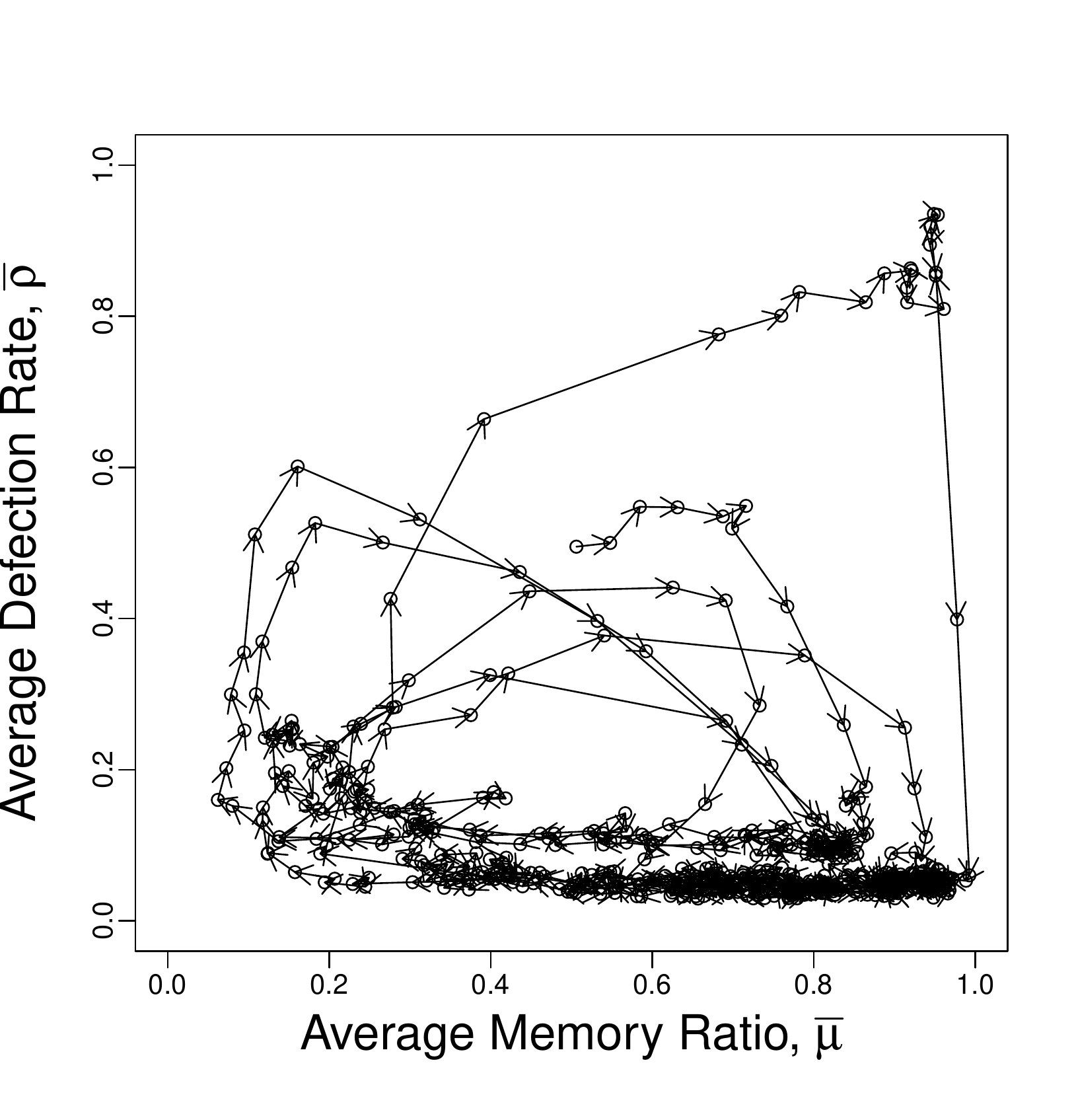}
	\caption{
		Co-evolution under aggressive defection for a single realization. 
		The same data, displayed in \reffig{fig:harmfulA}, graphed
		on a phase plane. 
		$(S,P,R,T) = (-7,-6,4,5)$.
	}
			\label{fig:harmfulB}
\end{figure}

We displayed the same data on the phase plane in \reffig{fig:harmfulB}.
Under aggressive defection, defectors with lower memory size, have no chance to survive. 
Thus, the average genotype of the population moves towards a point close to 
($\overline{\mu}$,$\overline{\rho}$)$\rightarrow (1,1)$
on the phase plane.
That point corresponds to a defective population with high memory size. 
Population genotype can stuck around this point or 
it can escape to continue its trending cyclic behavior. 
That depends not only the payoff matrix but also 
the dynamic composition of the heterogeneous population at any given time. 
In the next section, we will try to explore the effect of payoff matrix structure on the co-evolution of memory size and cooperation.

\section{The effect of payoff matrix structure}
\label{sec:Payoff}

Each payoff matrix has its own dynamics and it is very hard to make generalizations. 
We need to identify correctly the principal driving forces in our model,
and how they will affect the co-evolution of memory and cooperation.
To this end, we will reformulate the payoff matrix.
Note that the payoff matrices given 
in \reffig{fig:payoffMatrixDonation} and \reffig{fig:payoffMatrixEpstein} 
have some common properties.
Within both matrices the column differences are equal, i.e., $R - S = T-P$.
The row differences are also kept equal, i.e., $T - R  = P - S$.
If we generalize these differences 
we obtain two factors that are critical in the dynamics: 
(i)~how much it is dangerous to receive a defection, i.e. the column differences, and 
(ii)~how much it is tempting to defect, i.e., the row differences.
Thus we introduce the following 
two principal factors of threat and greed.

\begin{figure}[t]
\centering
	\includegraphics[scale=0.7]{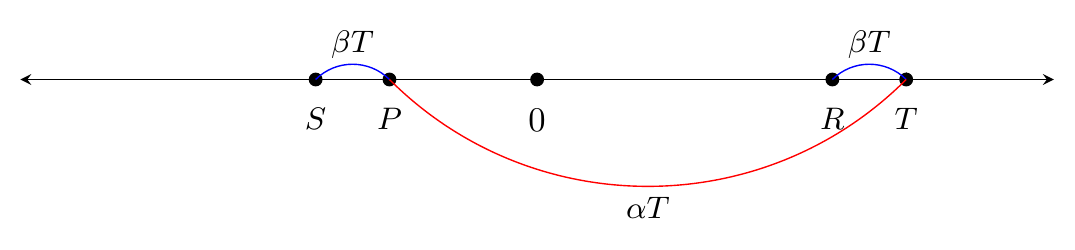}
	\caption{
		The visual representation of payoffs on a number line. 
		Note the fact that 
		$\alpha > 1$ makes  $S < P < 0$ and $\beta > 1$ makes $R < 0$.
	}
	\label{fig:numberLine}
\end{figure}

\begin{itemize}
	\item 
	\textbf{Threat factor, $\alpha$.}
	How to measure the difference between receiving a cooperation 
	and receiving a defection?
	The answer can be found in the payoff matrix seen in \reffig{fig:payoffMatrixSPRT}.
	For an agent that chooses to cooperate,
	the difference between receiving a cooperation and 
	receiving a defection is given by $R-S$.
	For an agent that chooses to defect,
	it is given by $T-P$.
	For simplification, 
	consider the case of $ R - S = T-P =\alpha T$
	where $T>0$ and $\alpha > 0$,
	as it is shown in \reffig{fig:numberLine}.
	Now irrespective of the chosen action, 
	receiving a defection 
	causes an extra cost of $\alpha T$ 
	in terms of payoffs
	when it is compared to receiving a cooperation.
	It is worth to emphasize that $\alpha  = 1$ is a critical value.
	For $\alpha  < 1$,
	$P$ is positive.
	At $\alpha  = 1$, $P$ becomes equal to $0$ 
	and for $\alpha > 1$,
	we have $P < 0$.
	Thus  $\alpha > 1$ corresponds to the case of aggressive defection ($S < P <0$).
	Hence we call $\alpha$ as the \emph{threat factor}.
	For $\alpha > 1$, increasing $\alpha$ means increasing level of threat.
	On the other hand, $\alpha \le 1$ means absence of threat.

	\item 
		\textbf{Greed factor, $\beta$.}
		How to measure the difference between taking the two actions of 
		defection and cooperation? 
		When the opponent is cooperating,
		the difference between defecting and cooperating is given by $T - R$.
		When the opponent is defecting,
		the same difference is given by $P - S$.
		Again for simplification
		consider the case of  $T - R = P - S = \beta T$
		where $T > 0$ and $\beta > 0$, 
		as it is shown in \reffig{fig:numberLine}.
		Now
		irrespective of the opponent's action,
		choosing to defect makes an extra benefit of $\beta T$ in terms of payoffs 
		when it is compared to choosing to cooperate.
		Thus, we call  $\beta$ as the \emph{greed factor}.
		When $\beta = 0$, playing cooperation or defection makes no difference.
		But whenever $\beta$ gets larger, defection becomes more tempting.
Since the case of $\beta > 1$ makes $R<0$, it turns out to be uninteresting.
This is because, when mutual cooperation payoff $R$ is also negative,
there will be no motivation for choosing to cooperate.
\end{itemize}
As an interpretation,  
choosing to defect brings an extra benefit of $\beta T$ 
(the row differences in the payoff matrix) 
and 
receiving a defection causes an extra cost of $\alpha T$ 
(the column differences).

\subsection{Threat Game}
Starting with a fixed positive value for $T$,
we can rewrite $S$, $P$, and $R$ in terms $\alpha$, $\beta$ and $T$.
That makes
$S = (1 - \alpha - \beta) T$,
$P = (1 - \alpha) T$, and
$R = (1 - \beta) T$.
IPD condition of $S < P < R < T$ implies that
$0 < \beta < \alpha$.
Since all payoff values are multiples of $T$,
the score of any agent will be also a factor of $T$.
When the normalized score is calculated, $T$ factor cancels out and 
we have expression in terms of $\alpha$ and $\beta$ only. 
Therefore, 
without loss of generality, 
we can take $T = 1$.
Our extensive simulations for different values of $T \in \{ 5, 50,100 \}$ have confirmed that 
the dynamics are not dependent on $T$.
Finally, we have the normalized payoff matrix given in \reffig{fig:payoffMatrixAlphaBetaNormalized},
which has only two parameters,
namely, 
the threat factor $\alpha$ and 
the greed factor $\beta$.
We call this special form of the IPD game 
as \emph{Threat game}.

This family of payoff matrices is a special case of all possible payoff matrices,
yet its is an important generalization which covers 
the donation game and also 
matrices that Epstein used.
(i)~The payoff matrix structure of the Donation game,
given in \reffig{fig:payoffMatrixDonation},
can be thought as a subset of Threat game for 
$b$
and 
$c$ 
as
column and 
row differences, 
respectively. 
The Donation game lies on the line segment of $\alpha = 1$ and $0<\beta  = \frac{c}{b} < 1$
in the $\alpha$-$\beta$ plain
in \reffig{fig:AlphaBeta}.
(ii)~Likewise, the payoff matrix structures,
given in \reffig{fig:payoffMatrixEpstein}, 
used by Epstein in Ref~\cite{%
	Epstein},
can be represented by
$T + R$ and $T - R$ 
as column and row differences,
respectively.
Hence
they correspond to the points on the line segment of $\alpha + \beta = 2$ again for $0<\beta < 1$.

\subsection{Observations}

\begin{figure*}[t]
	\begin{subfigure}[b]{\textwidth}
	\hspace*{-0.5cm}
		\includegraphics[scale=0.8]{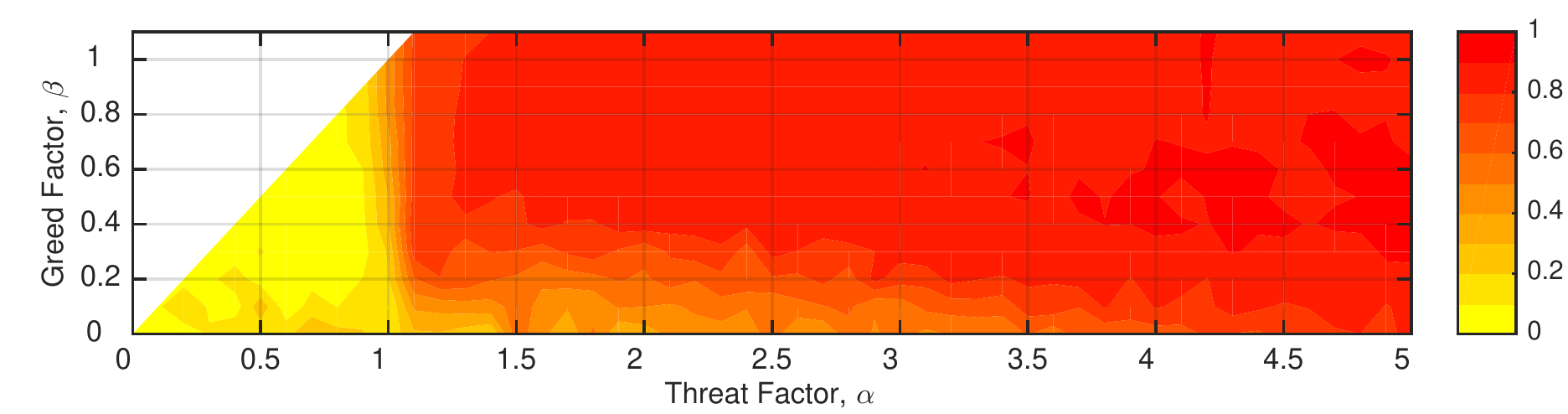}
		\caption{
			Average memory ratio $\overline{\mu}$ as 
			a function of $\alpha$ and $\beta$
		}
		\label{fig:AlphaBetaMu}
	\end{subfigure}\\
	\begin{subfigure}[b]{\textwidth}
	\hspace*{-0.5cm}
		\includegraphics[scale=0.8]{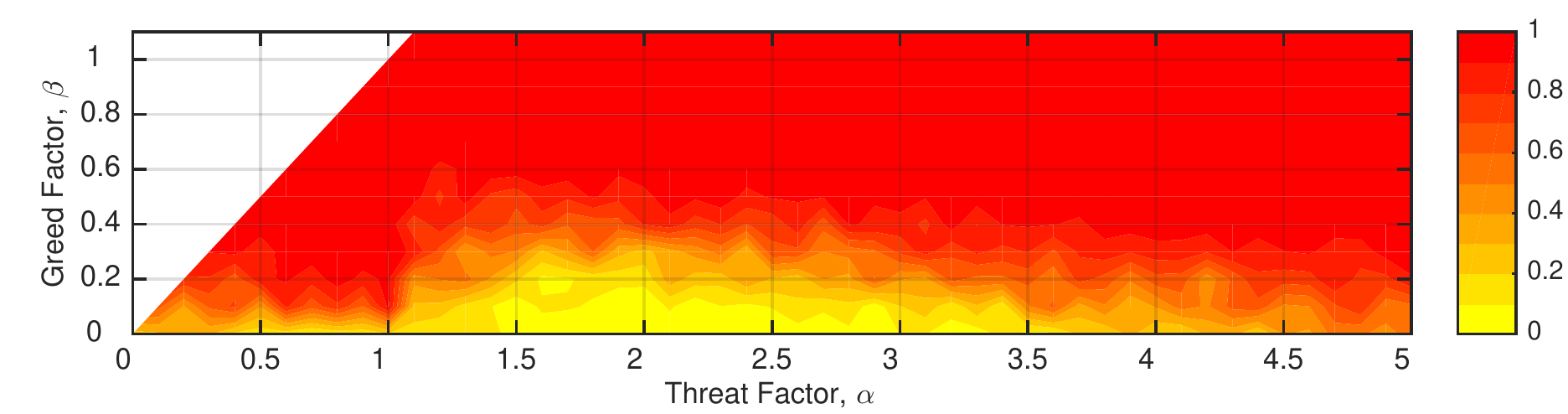}
		\caption{
			Average defection rate $\overline{\rho}$ as 
			a function of $\alpha$ and $\beta$
		}
		\label{fig:AlphaBetaRho}
	\end{subfigure}
	\caption{
		The effect of increasing level of threat and greed factors
		on the co-evolutionary dynamics of memory size and cooperation.
		The cases for $\alpha < \beta$ are omitted since they do not fulfill the conditions of the IPD game.
	}
	\label{fig:AlphaBeta}
\end{figure*}

We investigate the effect of increasing level of threat and greed factors
on the co-evolutionary dynamics of memory size and cooperation for $T = 1$.
\reffig{fig:AlphaBeta} visualizes 
how $\overline{\mu} $ and $\overline{\rho}$ change 
as a function of ($\alpha$, $\beta$) pairs.
In both figures of \reffig{fig:AlphaBetaMu} and \reffig{fig:AlphaBetaRho},
$x$-axis is the threat factor $\alpha  \in [0,5]$ and 
$y$-axis is the greed factor $\beta  \in [0,1.1]$.
Both $\alpha$ and $\beta$ have incremental steps of $0.1$ in their given ranges.
We omitted the case of $\alpha < \beta$ in \reffig{fig:AlphaBeta} since 
IPD condition of $S < P < R < T$ implies that
$0 < \beta < \alpha$.
But we showed the results for $\alpha = \beta$ and $\beta=0$ to see the limiting conditions.
The values of $\overline{\mu} $ and $\overline{\rho}$
are the averages over $20$ realizations.
We have only considered the average genotypes of the  
last $100$ subsequent generations.
That is from generation number $400$ to $500$.

\subsubsection{Evolution of memory}

In \reffig{fig:AlphaBetaMu}, we show the effect of threat and greed factors on the evolution of memory size. 
For the evolution of memory, $\alpha = 1$ is critical. 
$P$ becomes negative for $\alpha>1$, as it is shown in \reffig{fig:numberLine}.
So the change of $\alpha$ value from smaller than $1$ to greater than $1$
corresponds to the change from absence of threat to presence of threat.
In \reffig{fig:AlphaBetaMu},
we observe that 
the average memory size exhibits a major transformation from its lowest values 
to its highest values
when $\alpha$ becomes greater than $1$.
This shows clearly how threat fosters greater memory size.
We observe no direct impact of greed factor on the memory size.

\subsubsection{Evolution of cooperation}

In \reffig{fig:AlphaBetaRho}, we show the effect of threat and greed factors on the evolution of cooperation. 
For $\beta = 0$, agents have no preference neither for defection nor for cooperation. 
Cooperation may succeed for $\beta = 0$.
But it does not fulfill the conditions of IPD.
As greed factor $\beta$ gets larger, 
obviously it becomes harder to resist to the temptation to defect and 
cooperation vanishes totally for $\beta > 0.5$.
This is also true even if agents have high memory size.
It seems that threat has a non-linear effect on cooperation.
To better understand the underlying dynamics, we should reconsider the interdependence between 
threat, memory and cooperation.
Memory has a positive effect on cooperation,
but cooperation has a negative effect on memory.
\begin{itemize}

	\item 
	The role of memory is to block interactions with agents that are perceived as defectors.
	The increase in memory size can be thought as an introduction of (conceptual) barriers against 
	interaction with defectors.
	When memory size grows, defectors incur isolation. 
	Defectors can not gather enough fitness values for reproduction and 
	they are eliminated.
	
	\item 
	High memory size surprisingly raises the risk of self-isolation,
	especially in a population mostly composed of coopeartors.
	Cooperators can be perceived as defectors due to the small number of interactions
	in which they happened to defect more than cooperate.
	As a result of misperceptions,
	cooperators with high memory size can refuse to interact with other cooperators,
	hence they end up with lower fitness values.
	In the subsequent generations, cooperators get rid of their memory.
\end{itemize}
Threat calls for high memory size.
\begin{itemize}

	\item
	In the absence of threat, $\alpha \leq 1$, 
	receiving a defection brings non-negative payoffs, $0 \leq S<P$.
	In this case, refusing a defector, 
	due to the high memory size, 
	turns out to be disadvantageous. 
	So agents with high memory size are again eliminated throughout the evolution.
	
	\item 
	In the presence of threat, $\alpha > 1$, 
	receiving a defection brings negative payoffs, $S < P < 0$.
	This time, 
	refusing a defector and having a high memory size become advantageous.
	So average memory size has a tendency to increase in the presence of threat.
\end{itemize}
Increasing level of threat, primarily increases the memory size,
which, in turn, fosters cooperation. But further increase in the level of threat, can not help cooperation. 
In the next section we discuss the limits to cooperation.

\subsubsection{Drawing the boundaries of cooperation}
\label{sec:bounds}
We have made further simulations for greater number of generations.
We have observed that
an increase in the number of generations, up to a certain point,
has a positive effect on cooperation.
This raises the question whether there is a limit to cooperation.
So we decided to make an analytical effort
to draw the boundaries of cooperation, 
by simplifying our model as much as possible for the benefit of cooperators.
Let's think that there exists only 
pure cooperators, that always cooperate, and pure defectors, that always defect, in the population.
In order to favor cooperators, suppose 
memory ratio $\mu_i = 1$
for all agents.
So agents will remember the past actions of all their opponents.
Suppose also mutation is prohibited.
Hence,
both cooperators and defectors will always be cautious against other defectors,
throughout the evolutionary process.
Since agents have enough memory
they will play utmost one round with one particular defector
and refuse to play with it afterwards.

Suppose there are $d \in [0,N]$ defectors in the population.
In our model, each pair of agents are matched $\tau$ times on the average.
So a cooperator will play $\tau$ times with $(N-d-1)$ other cooperators 
(and receive the reward payoff $1-\beta$)
but it will play only once with $d$ defectors (and receive the sucker payoff $1-\alpha-\beta$).
Hence, the average performance of a particular cooperator equals to the following.
\[
P_C =  \tau (N-d-1) (1-\beta) + d (1-\alpha-\beta)
\]
The average performance of a particular defector
can be obtained in a similar fashion.
A defector will play only once with $(N-d)$ cooperators (and receive the temptation payoff $1$)
and will again play only once with ($d-1$) other defectors (and receive the punishment payoff $1-\alpha$).
\[
	P_D = (N-d) + (d-1)(1-\alpha).
\]
In order to have $P_C > P_D$, $\alpha$ should satisfy
\begin{align}
 	\alpha <  (N-d-1)(\tau(1-\beta)-1) -d\beta.
	\label{eq:alpha}
\end{align}
Extremely beneficial conditions for cooperation can be achieved 
by setting $\beta = 0$. 
Even in this condition,
population can resist to a single ($d=1$) defector, up to a certain point.
By setting $d=1$ and $\beta = 0$ in \refeq{eq:alpha},
we obtain $\alpha_{1}$ and $\alpha_{2}$ given as
\[
	\alpha < \alpha_{1} = (N-2) (\tau-1) < N \tau = \alpha_{2}.
\]
For greater values of $\alpha > \alpha_{2}$,
irrespective of the generation number and the greed factor $\beta$, 
it becomes impossible to resist defectors.
%
Absence of greed ($\beta=0$) makes cooperation and defection indifferent
in terms of payoffs, see \reffig{fig:payoffMatrixAlphaBetaNormalized}.
Then why pure cooperators ($\rho = 0$) fail and pure defectors ($\rho = 1$) rise for 
$\alpha >  \alpha_{2}$?
This is simply because a defector can not receive a defection from itself.
So cooperators receive $1$ more defection than defectors does.
This difference becomes impossible to compensate when the level of threat $\alpha$ 
is greater than the average number of rounds $\tau$ with each opponent, multiplied by number of agents $N$, multiplied by the maximum payoff per round $T=1$.
%
%
This was an excessive simplification to show that there exists a limit to cooperation.

Our bounds for $\alpha$ are not tight.
The $\alpha_{2}$ value is $3000$ for $\tau = 30$ and $N = 100$.
We test the bounds by simulations, 
in which populations are not composed of only 
pure cooperators and pure defectors with memory ratio of $1$
but of heterogenous agents with various 
defection rates
and memory ratios.
We obtained much more modest values of $\alpha_{2}$.
We have obtained 
$\alpha_{2} = 10$  for generation number $500$,
and 
$\alpha_{2} = 20$  for generation number $2500$.
Above these $\alpha_{2}$ values we do not observe any cooperation.
%

\subsubsection{Co-evolution of memory and cooperation}
Necessity is the mother of invention. 
We see a positive function of threat in having a greater memory size. 
On the other hand,
greater memory size raises the risk of self-isolation.
Threat and misperceptions among cooperators surprisingly cause a second source of dilemma on the memory size. 
Thus not only cooperation but also memory size constitutes a dilemma in our model.
We can summarize the resulting dynamics of our model in three predominant categories. 
\begin{enumerate}
	
	\item[(i)] 
	In the absence of threat ($\alpha \leq 1$), greater memory size is unfavorable to evolutionary success.
	And cooperation collapses without memory.
	Thus, the average genotype of the population moves to a point close to 
	$(\overline{\mu}, \overline{\rho}) = (0, 1)$.
	\reffig{fig:PhasePlaneAxelrod} shows the triumph of memoryless defection over time, 
	when there is no threat.
	
	\item[(ii)]  
	In the presence of an appropriate level of threat ($1 < \alpha < \alpha_{2}$)
	and under low greed factor ($\beta < 0.5$),
	the trajectory of the evolving population can
	exhibit emergent oscillatory dynamics. 
	Memory size acts like an immune response of the subsequent generations. 
	\reffig{fig:harmfulA} helps us to visualize the emerging oscillatory dynamics 
	for a single realization, in the presence of an appropriate level of threat. 
	Memory crashes when defection rate is low, 
	but spikes up as a response to growing defection rate in the generation, 
	then crashes again and recover again depending on the average defection rate of the 
	subsequent generations. 
	Not individual agents, 
	but populations from generations to generations evolved to develop 
	some kind of protection mechanism against aggressive defection. 
	This
	is an un-programmed functionality that emerged via evolutionary dynamics.
	
	\item[(iii)] 
	In the presence of threat ($\alpha > 1$) and for high greed factor ($\beta > 0.5$)
	the average genotype of the population moves towards a point close to 
	$(\overline{\mu}, \overline{\rho}) = (1, 1)$
	which corresponds to a defective population with high memory size. 
	This is also true in the presence of an extreme threat ($\alpha >  \alpha_{2}$). 
	As $\alpha$ gets larger and approaches to $\alpha_{2}$, defection starts to cause an extreme damage and
	it gets harder for cooperators to resist.
\end{enumerate}
To our conclusion,
the dose of the threat makes the resistance for cooperation,
especially when the greed factor is low.
Lastly, to understand the impact of attention,
we compared the dynamics under selective attention (forget preferentially cooperators)
and the dynamics without attention (forget at random without preference).
It seems like attention favors cooperation and 
disfavors defection, especially for moderate values of threat and greed factors.
Attention can only make a difference when memory size is limited.
For higher threat and greed factors
memory size gets very close to its maximum value 
and memory becomes sufficient to remember all opponents
that, in turn, makes attention less critical.
Nevertheless, we think that the effect of attention on the co-evolutionary dynamics of memory and cooperation 
requires a research of its own.
We leave it as a future work.

\section{Conclusion}
\label{sec:Conclusion}
	In the research of cooperation,
	the effect of negative mutual punishment payoff,  $P<0$, is usually omitted,
	as in the case of the Donation Game.
	Yet there are some studies with negative payoffs for receiving a defection. 
Our model shares with Epstein, 
the effect of negative payoffs in the emergence of cooperation~\cite{Epstein}.
But it differs in many other respects.
First,
our model allows agents to have varying memory size and defection rate,
whereas Epstein's model allows only pure cooperators 
and pure defectors
with zero-memory.
The structure of a system,
determines \emph{who interacts with whom}
and causes its dynamic behavior.
In our model,
agents can select their partners and memory has a critical role, 
as it is used to hinder interactions with defectors.
In \refcite{Epstein},
it is only the spatial aspects of the environment which 
hinders cooperators from interacting with defectors.
In other words,
Epstein used physical barriers to avoid interactions with defectors,
while we have used conceptual barriers for it. 
Instead of studying cooperation under fixed payoff matrix structure,
we proposed to reformulate the payoff matrix structure of IPD game by introducing 
threat and greed factors,
and showed their effect on the co-evolution of memory and cooperation.

We observe that the greater memory size is unfavorable to evolutionary success when there is no
threat.
One can find this, initially, deeply counterintuitive and not realistic. 
But there are cases where species lost their brains as a result of evolution. 
According to Frank Hirth, in their ancient evolutionary past, sea sponges did have neurons~\cite{Hirth2010}.
Some extremely simple animals, such as sea squirts, simplify their brains during their lifetimes. 
Sea squirt has a nervous system in order to navigate in the sea. 
Its only goal is to find a suitable rock to live on for the rest of its life. 
When it implants on a rock, the first thing it does is to digest its nervous system. 
Without a problem to solve, there is no need to waste energy on a brain. 
In order for evolution to promote increased brain size, its benefits, e.g. against predation threat, 
must outweigh the high energetic costs~\cite{Kotrschal}.
One of the most striking example related to the evolution of brain size belongs to humans. In the past 20.000 years, the human brain has shrunk by about the size of a tennis ball~\cite{DomesticatedBrain2014}.
Nobody knows exactly why. According to a leading theory, the incredible decline in human brain size is a by-product of domesticity. The shift from the threatening lifestyle of hunter-gatherers to the highly cooperative and more secure lifestyle of agricultural society has led to the reduction in brain size. Our results support this theory.

We have shown the positive effect of an appropriate level of threat in having a greater memory size
which, in turn, favors cooperation.
This finding is in parallel with
other forms of delicate balance
(for the level of environmental harshness~\cite{szolnoki2014binary}  
and for the level of punishment fines~\cite{helbing2010punish})
within which cooperation thrives best in 
spatial evolutionary games.
It is possible to make analogies with two different scientific results from immunology and experimental psychology.
It is thought that the immune system functions by making distinction between self and non-self. 
This viewpoint is renewed with the idea 
that the immune system is more concerned with entities that do damage than with those that are foreign~\cite{Matzinger12042002}.
Actually, threat calls for taking countermeasures against the would-be-exploiter. 
Experimental evidence from psychology has shown that 
the cooperation typically collapses in the absence of sanctioning possibilities~\cite{Gurerk108}.
The threat of punishment is the key to maintain and promote cooperation~\cite{Camerer47,Jordan2016}. 
To conclude, in order for cooperation to emerge, selfish beings need to be exposed to an appropriate level of threat. 
When defection is harmless agents tend to defect and when defection cause an extreme damage, 
cooperators have no chance to survive. We observe that the conditions for the emergence of cooperation are very subtle. 
To increase the immunity of cooperation, differentiation of cooperators or some kind of collective memory 
can be incorporated to our model as a future work.

\section*{Acknowledgments}

This work was partially supported 
by Bogazici University Research Fund, BAP-2008-08A105, 
by the Turkish State Planning Organization (DPT) TAM Project, 2007K120610, 
and 
by COST action MP0801.

\bibliography{TheDose}

\end{document}